\documentstyle[12pt]{article}

\begin{document}
\title{Current-Carrying Cosmic Strings in Scalar-Tensor Gravities}
\author{C. N. Ferreira$^1$\thanks{crisnfer@cat.cbpf.br} ,  
M. E. X. Guimar\~aes$^2$\thanks{emilia@mat.unb.br} and  J. A. 
Helay\"el-Neto$^1$\thanks{Also at Grupo de F\'{\i}sica Te\'orica, 
Universidade Cat\'olica de Petr\'opolis. helayel@cat.cbpf.br}\\
\mbox{\small{1. Centro Brasileiro de Pesquisas F\'{\i}sicas, 
Rua Dr. Xavier Sigaud, 150. Urca}} \\ 
\mbox{\small{CEP: 22290-180. Rio de Janeiro - RJ, Brazil}} \\
\mbox{\small{2. Universidade de Bras\'{\i}lia, Departamento de Matem\'atica}} \\
\mbox{\small{Campus Universit\'ario - CEP: 70910-900. Bras\'{\i}lia - DF, Brazil}}} 
\maketitle
\begin{abstract}
We study the modifications on the metric of an isolated self-gravitating 
bosonic superconducting cosmic string in a scalar-tensor gravity in 
the weak-field approximation. These modifications are 
induced by an arbitrary coupling of a massless scalar field 
to the usual tensorial field in the gravitational Lagrangian. 
The metric is derived by means of a matching at the string radius 
with a most general static and cylindrically symmetric solution of the 
Einstein-Maxwell-scalar field equations. We show that this metric depends on 
five parameters which are related 
to the string's internal structure and to the solution of the 
scalar field.  We compare our results with those obtained in the 
framework of General Relativity. 
\end{abstract}

\section{Introduction}
The assumption that gravity may be intermediated by a scalar field 
(or, more generally, by many scalar fields) in addition 
to the usual symmmetric rank-2 tensor has considerably revived in the recent years. 
From the theoretical point of view, they seem to be the most natural 
alternative to General Relativity. Indeed, most attempts 
to unify gravity with the other interactions predict the existence of 
one (or many) scalar(s) field(s) with gravitational-strength couplings. 
If gravity is essentially scalar-tensorial, there will be direct 
implications for cosmology and experimental tests of the gravitational interaction (we refer the 
reader to Damour's recent account on 
``Experimental Tests of GR" \cite{dam}). 
In particular, any gravitational
phenomena will be affected by the variation of the gravitational ``constant" $\tilde{G}_{0}$. At 
sufficiently high energy scales where gravity becomes 
scalar-tensor in nature \cite{green}, it seems worthwhile to analyse the behaviour of 
matter in the presence of a scalar-tensorial gravitatinal field, specially those 
which originated in the early universe, such as cosmic strings.  In this context, some a
uthors have studied solutions for cosmic 
strings and domain walls in Brans-Dicke \cite{rom}, in dilaton 
theory \cite{greg} and in more general scalar-tensor couplings 
\cite{mexg}. 

On the other hand, topological defects are expected to be formed 
during phase transitions in the early universe. Among them, cosmic strings have been 
widely studied in cosmology in connection with structure formation. In 1985, Witten  
showed that in many field theories cosmic strings behave as superconducting tubes and 
they may generate enormous currents of order $10^{20} A$ or more \cite{wit}. This fact 
has raised  interest to current-carrying strings and their eventual explanations 
to many astrophysical phenomena, such as origin of the primordial magnetic fields 
\cite{tan}, charged vaccum condensates \cite{nas} and sources of ultrahigh-energy 
cosmic rays \cite{hill}, among others. 

In ref. \cite{sen}, Sen have considered solutions of a superconducting string in the  
Brans-Dicke theory. The aim of this paper is to study the implications of a 
class of more general scalar-tensor gravities for a  
superconducting, bosonic cosmic string. 
In particular, we will be interested on the modifications 
induced on the string metric and their possible observable consequences 
on the current carried by the string. 
These modifications come from an arbitrary coupling of a massless scalar 
field to the tensor field in the gravitational Lagrangian. The action which 
describes these theories  (in the Jordan-Fierz frame) is 
\begin{equation}
{\cal S} = \frac{1}{16\pi} \int d^4x \sqrt{-\tilde{g}} 
\left[\tilde{\Phi}\tilde{R} - \frac{\omega(\tilde{\Phi})}{\tilde{\Phi}}
\tilde{g}^{\mu\nu}
\partial_{\mu}\tilde{\Phi}\partial_{\nu}\tilde{\Phi} \right] 
+ {\cal S}_{m}[\Psi_m, \tilde{g}_{\mu\nu}] , 
\end{equation}
where $\tilde{g}_{\mu\nu}$ is the physical metric in this frame, 
$\tilde{R}$ is the curvature scalar associated to it and 
${\cal S}_m$ denotes the action describing the general matter 
fields $\Psi_m$. These theories are metric, e.g., matter couples 
minimally and universally to $\tilde{g}_{\mu\nu}$ and not to 
$\tilde{\Phi}$. 

The main purpose of this paper is to study the influence of a scalar-tensorial 
coupling on the gravitational field of a current-carrying cosmic string described by 
Witten's model \cite{wit}. For this purpose, we need to solve the modified Einstein's 
equations having a current-carrying vortex as source of the spacetime. 
In General Relativity, the gravitational field of superconducting strings has been 
studied by many authors \cite{moss,ams,babul,lin,helli,pet2}. In particular, the 
following technics have b
een employed to derive the spacetime surrounding superconducting vortex: analytic 
integration of the Einstein's equations over the string's energy-momentum tensor 
\cite{moss}; linearization of the Einstein's equations using distribution's 
functions \cite{lin,pet2}; numerical integrations of the fields equations (Einstein plus material 
fields) \cite{ams}, among others. 

In this paper, we will make an adaptation of Linet's method \cite{lin} to our model. 
That is, we will solve the linearised (modified) Einstein's equations using 
distribution's functions while taking into account the scalar-tensor feature of gravity. 
This work is outlined as follows. In section 2, we describe the configuration of a  
superconducting string in scalar-tensor gravities. 
In section 3, we start by solving the equations for the exterior region. 
In the subsection 3.2, we solve the linearised equations by applying Linet's method, 
introduced in ref. \cite{lin}. Then, we match the exterior solution with
the internal parameters. In 3.3, we  derive the deficit angle associated to 
the metric found previously. We also compare our results with previous 
results obtained in the framework of General Relativity. 
Finally, in section 4, we end with some conclusions and discussions. 

\section{Superconducting String Configuration in Scalar-Tensor Gravities}

In what follows, we will search for a regular solution of a self-gravitating 
superconducting vortex in the framework of a scalar-tensor gravity. Hence, the simplest 
bosonic vortex arises from the action of the Abelian-Higgs 
$U(1) \times U'(1)$ model 
containing two pairs of complex scalar and gauge fields
\begin{eqnarray}
{\cal S}_m  & = & \int d^4x \sqrt{-\tilde{g}} \{ 
- \frac{1}{2}\tilde{g}^{\mu\nu}D_{\mu}\varphi D_{\nu}\varphi^* - \frac{1}{2}
\tilde{g}^{\mu\nu}D_{\mu}\sigma D_{\nu}\sigma^* \nonumber \\    
&& - \frac{1}{16\pi}\tilde{g}^{\mu\nu}\tilde{g}^{\alpha\beta}H_{\mu\alpha}H_{\nu\beta}
- \frac{1}{16\pi} \tilde{g}^{\mu\nu}\tilde{g}^{\alpha\beta}F_{\mu\alpha}
F_{\nu\beta} - V(\mid\varphi\mid , \mid\sigma\mid) \}
\end{eqnarray}
with $D_{\mu}\varphi \equiv (\partial_{\mu} + iqC_{\mu})\varphi$, 
$D_{\mu}\sigma \equiv (\partial_{\mu} + ieA_{\mu})\sigma$ and 
$F_{\mu\nu}$ and $H_{\mu\nu}$ are the field-strengths 
associated to the electromagnetic $A_{\mu}$ and gauge $C_{\mu}$ 
fields, respectively. The potential is  ``Higgs inspired" and 
contains appropriate $\varphi-\sigma$ interactions so that there occurs a 
spontaneous symmetry breaking
\begin{equation}
V(\mid\varphi\mid , \mid\sigma\mid) = \frac{\lambda_{\varphi}}{4}
(\mid\varphi\mid^2 - \eta^2)^2 + f\mid\varphi\mid^2\mid\sigma\mid^2 + \frac{\lambda_{\sigma}}{4}
\mid\sigma\mid^4 - \frac{m^2}{2}\mid\sigma\mid^2  ,
\end{equation}
with positive $\eta , f , \lambda_{\sigma}, \lambda_{\varphi}$ 
parameters. A vortex configuration arises 
when the $U(1)$ symmetry associated to the $(\varphi, C_{\mu})$ pair is 
spontaneously broken. The superconducting 
feature of this vortex is produced when the pair 
$(\sigma, A_{\mu})$, associated to the other $U'(1)$ symmetry 
of this model, is spontaneously broken in the core of the vortex. 

We restrict ourselves to contemplate configurations of an 
isolated and static vortex in the $z$-axis. In a cylindrical 
coordinate system $(t,r,\theta,z)$, such that $r \geq 0$ and 
$0 \leq \theta <2\pi$, we make the choice

\begin{equation}
\varphi = R(r)e^{i\theta} \;\; \mbox{and} \;\; 
 C_{\mu} = \frac{1}{q}[P(r) - 1]
\delta^{\theta}_{\mu} , 
\end{equation}
in much the same way as we proceed with ordinary (non-conducting) cosmic strings. 
The functions $R, P$ are functions of $r$ only. 
We also require that these functions  be regular everywhere 
and that they satisfy the usual boundary conditions for a 
vortex configuration \cite{niel}

\[
R(0) = 0 \;\; \mbox{and} \;\; P(0) =1 
\]
\begin{equation}
\lim_{r\rightarrow\infty} R(r)=\eta \;\; \mbox{and} 
\lim_{r \rightarrow \infty} P(r) = 0 .
\end{equation}

The $\sigma$-field is responsible for the bosonic current along 
the string, and the $A_{\mu}$ is the gauge field which produces 
an external magnetic field; their configuration are taken in 
the form

\begin{equation}
\sigma = \sigma(r)e^{i\psi(z)} \;\; \mbox{and} \;\;
A_{\mu} = \frac{1}{e}[A(r)- \frac{\partial{\psi}}{\partial z}]
\delta^{z}_{\mu}
\end{equation}
The pair $(\sigma,A_{\mu})$ is subjected to the 
following boundary conditions

\[
\frac{d\sigma(0)}{dr}=0 \;\; \mbox{and} \;\; A(0)=\frac{dA(0)}{dr}=0 
\]
\begin{equation}
\lim_{r\rightarrow \infty}\sigma (r) =0 \;\; \mbox{and} \;\; 
\lim_{r \rightarrow \infty}A(r) \neq 0.
\end{equation}
With this choice, we can see that $\sigma$ breaks electromagnetism 
inside the string and can form a charged scalar condensate in the string core. 
Outside the string, the $A_{\mu}$ field has a non-vanishing component along the 
$z$-axis which indicates that there will be a non-vanishing
energy-momentum tensor in the region exterior to the string. 

Although action (1) shows explicitly this gravity's scalar-tensorial character, 
for technical reasons, we choose to work in the 
conformal (Einstein) frame in which the kinematic terms of the scalar and the tensor  
fields do not mix 

\begin{equation}
{\cal S} = \frac{1}{16\pi G} \int d^4x \sqrt{-g} \left[ R - 
2g^{\mu\nu}\partial_{\mu}\phi\partial_{\nu}\phi \right] 
+ {\cal S}_{m}[\Psi_m,\Omega^2(\phi)g_{\mu\nu}] ,
\end{equation}
where $g_{\mu\nu}$ is a pure rank-2 tensor in the Einstein frame, 
$R$ is the curvature scalar associated to it and $\Omega(\phi)$ 
is an arbitrary function of the scalar field. 

Action (8) is 
obtained from (1) by a conformal transformation 
\begin{equation}
\tilde{g}_{\mu\nu} = \Omega^2(\phi)g_{\mu\nu} ,
\end{equation}
and by a redefinition of the quantity
\[
G\Omega^2(\phi) = \tilde{\Phi}^{-1}
\]
which makes evident the feature that any gravitational 
phenomena will be affected by the variation of the gravitation ``constant" $G$ 
in the scalar-tensorial gravity, and by introducing a new parameter 
\[
\alpha^2 \equiv \left( \frac{\partial \ln \Omega(\phi)}{\partial 
\phi} \right)^2 = [2\omega(\tilde{\Phi}) + 3]^{-1} , 
\]
which can be interpreted as the (field-dependent) coupling strength 
between matter and the scalar field. 
In order to make our calculations as broad as possible, we 
choose not to specify the factors $\Omega(\phi)$ and 
$\alpha (\phi)$ 
(the field-dependent coupling strength between matter and the scalar field), 
leaving them as arbitrary functions of the scalar field. 

In the conformal frame, the Einstein equations are modified.  A 
straightforward calculus shows that the ``Einstein" equations are 
\begin{eqnarray}
G_{\mu\nu} & = & 2\partial_{\mu}\phi\partial_{\nu}\phi -
g_{\mu\nu}g^{\alpha\beta}\partial_{\alpha}\phi\partial_{\beta}
\phi + 8\pi G T_{\mu\nu} \nonumber \\
\Box_{g}\phi & = & -4\pi G \alpha(\phi) T .
\end{eqnarray}
We note that the last equation brings a new information and shows that the 
matter distribution behaves as a source for $\phi$ 
and $g_{\mu\nu}$ as well. The energy-momentum tensor is defined 
as usual 
\begin{equation}
T_{\mu\nu} \equiv \frac{2}{\sqrt{-g}}\frac{\delta {\cal S}_m}
{\delta g_{\mu\nu}} ,
\end{equation}
but in the conformal frame it is no longer conserved 
$\nabla_{\mu}T^{\mu}_{\nu} = \alpha(\phi)T\nabla_{\nu}\phi$. It 
is clear from transformation (9) that we can relate quantities from both 
frames in such a way that $\tilde{T}^{\mu\nu} =
\Omega^{-6}(\phi)T^{\mu\nu}$ and $\tilde{T}^{\mu}_{\nu} =
\Omega^{-4}(\phi)T^{\mu}_{\nu}$. 

Guided by the symmetry of the source, we impose that the metric 
is static and cylindrically symmetric. We choose to work with a general 
cylindrically symmetric metric written in the form 
\begin{equation}
ds^2 = e^{2(\gamma -\Psi)}(-dt^2 + dr^2) + \beta^2 e^{-2\Psi} 
d\theta^2 + e^{2\Psi}dz^2 ,
\end{equation}
where the metric functions $\gamma,\Psi,$ and $\beta$ are functions of 
$r$ only. In addition,  the metric functions 
satisfy the regularity conditions at the axis of symmetry $r=0$
\begin{equation}
\gamma = 0,  \;\; \Psi =0, \;\; \frac{d\gamma}{dr}=0, \;\; 
\frac{d\Psi}{dr} =0,  \;\; \mbox{and} \;\; \frac{d\beta}{dr} =0 .
\end{equation}

With metric given by expression (12) we are in a position to write the full 
equations of motion for the self-gravitating superconducting vortex in 
scalar-tensorial gravity. In the conformal frame, these equations are 
\begin{eqnarray}
\beta^{''} & = & 8\pi G\beta e^{2(\gamma - \Psi)} [T^{t}_{t} + T^{r}_{r}] \nonumber \\
(\beta\Psi^{'})^{'} & = & 4\pi G\beta e^{2(\gamma - \Psi)} [T^{t}_{t} + T^{r}_{r}
+T^{\theta}_{\theta} - T^{z}_{z}] \nonumber \\
\beta^{'}\gamma^{'} & = & \beta(\Psi^{'})^{2} - \beta(\phi^{'})^{2} + 
8\pi G e^{2(\gamma - \Psi)}T^{r}_{r} \nonumber \\
(\beta\phi^{'})^{'} & = & - 4\pi G \alpha(\phi)\beta e^{2(\gamma - \Psi)} T  ,
\end{eqnarray}
where $(')$ denotes ``derivative with respect to r". 
The non-vanishing components of the energy-momentum tensor (computed using equation (11)) are 
\begin{eqnarray}
T^{t}_{t} & = &  - \frac{1}{2}\Omega^{2}(\phi) \{ 
e^{2(\Psi - \gamma)}(R'^{2} + \sigma'^{2}) + \frac{e^{2\Psi}}
{\beta^2}R^2P^2 + e^{-2\Psi}\sigma^2A^2  \nonumber \\
& & + \Omega^{-2}(\phi)e^{-2\gamma}(\frac{A'^2}{4\pi e^2}) + 
\Omega^{-2}(\phi)\frac{e^{2(2\Psi - \gamma)}}{\beta^2}(\frac{P'^2}
{4\pi q^2}) + 2\Omega^2(\phi)V(R,\sigma) \} \nonumber \\ 
T^{r}_{r} & = &  \frac{1}{2}\Omega^2(\phi) \{ e^{2(\Psi - \gamma)}
(R'^2 + \sigma'^2) - \frac{e^{2\Psi}}{\beta^2} R^2P^2 - e^{-2\Psi}\sigma^2A^2 \nonumber \\
& & + \Omega^{-2}(\phi)e^{-2\gamma}(\frac{A'^2}{4\pi e^2}) + \Omega^{-2}(\phi)
\frac{e^{2(2\Psi - \gamma)}}{\beta^2}(\frac{P'^2}{4\pi q^2}) - 2\Omega^2(\phi)
V(R,\sigma) \} \nonumber \\
T^{\theta}_{\theta} & =  & - \frac{1}{2}\Omega^2(\phi) \{ e^{2(\Psi - \gamma)}
(R'^2 + \sigma'^2) - \frac{e^{2\Psi}}{\beta^2} R^2P^2 + 
e^{-2\Psi}\sigma^2A^2  \\
& & + \Omega^{-2}(\phi)e^{-2\gamma}(\frac{A'^2}{4\pi e^2}) - \Omega^{-2}(\phi) 
\frac{e^{2(2\Psi - \gamma)}}{\beta^2}(\frac{P'^2}{4\pi q^2}) + 2\Omega^2(\phi)
V(R,\sigma) \} \nonumber \\
T^{z}_{z} & = & - \frac{1}{2}\Omega^2(\phi) \{ e^{2(\Psi - \gamma)} 
(R'^2 + \sigma^2) + \frac{e^2\Psi}{\beta^2}R^2P^2 - e^{-2\Psi}\sigma^2
A^2 \nonumber \\
& & - \Omega^{-2}(\Phi)e^{-2\gamma}(\frac{A'^2}{4\pi e^2}) + \Omega^{-2}(\phi)
\frac{e^{2(2\Psi -\gamma)}}{\beta^2}(\frac{P'^2}{4\pi q^2}) +  2\Omega^2(\phi)
V(R,\sigma) \nonumber \}
\end{eqnarray}

As we said before, the energy-momentum tensor is not conserved in the conformal frame. 
Instead, the equation
\[
\nabla_{\mu}T^{\mu}_{\nu} = \alpha(\phi) T \nabla_{\nu}\phi ,
\]
where $T$ is the trace of the energy-momentum tensor, gives an 
additional relation between the scalar field $\phi$ and 
the source.

In the next section, we will attempt to solve the field equations (14). 
For the purpose of these calculations, we can divide the space into two regions: 
an exterior region $r > r_0$, where all the fields drop away rapidly and the 
only survivor is the magnetic field; and an 
interior region $r \leq r_0$, where all the string's field 
contribute to the energy-momentum tensor. Conveniently, $r_0$ has the same 
order of magnitude of the string radius. Then, we match the exterior and the 
interior solutions (to first order in 
$\tilde{G}_{0} = G\Omega^2(\phi_0)$, where $\phi_0$ is a constant) 
providing a relationship between the internal 
parameters of the string and the spacetime geometry.

\section{Superconducting String Solution in Scalar-Tensor Gravities}

\subsection{The Exterior Solution and the Modified Rainich Algebra:}

In this region, $r > r_0$, the electromagnetic field is the only field which 
contributes to the energy-momentum tensor. Therefore, 
the energy-momentum tensor has the form\footnote{Just as a reminder, 
throughout this paper we will work in the conformal frame for the sake of simplicity. 
Also, for convenience, we work in units such that $\hbar = c=1 $ and keep Newton's ``constant"  
$G$.} 

\begin{equation}
T^{\mu\nu} = \frac{1}{4\pi} \left[ F^{\mu\alpha}F^{\nu}_{\alpha} - 
\frac{1}{4}g^{\mu\nu}F^{\alpha\beta}F_{\alpha\beta} \right]
\end{equation}
with the following algebraic properties

\[
T^{\mu}_{\mu} = 0 \;\;\; \mbox{and} \;\;\; 
T^{\alpha}_{\nu}T^{\mu}_{\alpha}=\frac{1}{4}\delta^{\mu}_{\nu}(T_{\alpha\beta}T^{\alpha\beta}).
\]
which leads expression (15)  to take a simple form

\begin{equation}
T^t_t = - T^r_r= T^{\theta}_{\theta}= - T^z_z = 
- \frac{1}{2}e^{-2\gamma} \left( \frac{A'^2}{4\pi e^2} \right) .
\end{equation}
Thus, our problem is reduced to solve the modified Einstein's equations 
with source given by (16). That is,

\begin{eqnarray}
\beta'' & = & 0 \nonumber \\
(\beta\Psi')' & = & 4\pi G\beta e^{2(\gamma - \Psi)} [T^t_t - T^z_z] \nonumber \\
\beta'\gamma' & = & 8\pi G\beta e^{2(\gamma - \Psi)} T^r_r +
\beta(\Psi')^2 - \beta (\phi')^2 \nonumber \\
(\beta\phi')'  & = & 0 \; . 
\end{eqnarray} 

In General Relativity (i.e, in the absence of the dilaton field), 
these equations have been previously investigated by many authors \cite{louis}. 
A source of the form (16) leads to some algebraic conditions on the curvature scalar 
and the Ricci tensor, 
known as the Rainich conditions:
\[
R \equiv R^t_t + R^r_r + R^{\theta}_{\theta} + R^z_z = 0 ,
\]
and
\[
(R^t_t)^2 = (R^r_r)^2 = (R^{\theta}_{\theta})^2 = (R^z_z)^2 .
\]
The two equations above admit three sets of solutions: the magnetic case, 
the electric case and a third case which can correspond to either a static 
electric or a static magnetic field 
aligned along the $z$-axis \cite{louis}. The superconducting string 
defined in Witten's model (2) correponds to the magnetic case:
\begin{equation}
R^t_t= R^{\theta}_{\theta} \;\;\; R^{\theta}_{\theta} = R^z_z \;\; 
\mbox{and} \;\;  R^t_t = - R^r_r .
\end{equation} 

In a scalar-tensor gravity, we notice however that the Rainich conditions are no longer valid 
because of the very nature of the modified Einstein's equations (actually an 
Einstein-Maxwell-dilaton system). Instead of the algebraic conditions stated above, we have now:
\begin{equation}
R \equiv R^t_t + R^r_r + R^{\theta}_{\theta} + R^z_z = 
2(\phi')^2 e^{2(\Psi - \gamma)} ,
\end{equation}
and the analogous to the magnetic case in the scalar-tensor gravity is a solution of the form:
\begin{equation}
R^t_t = R^{\theta}_{\theta} \;\;\; R^{\theta}_{\theta} = - R^z_z 
\;\; \mbox{and} \;\; R^t_t = - R^r_r - 2(\phi')^2 e^{2(\Psi - \gamma)}.
\end{equation}

We are now in a position to solve the modified Einstein's equations. 
The first and last equations in (18) can be solved straightforwardly:
\begin{eqnarray}
\beta(r) & = & B r  \nonumber \\
\phi(r) & = & l\ln(r/r_0) .
\end{eqnarray}
The second and third equations in (18) are solved with the help 
of the algebraic conditions (20) and (21):
\[
  \gamma'' +\frac{1}{r} \gamma' = 0 , \]
\[
\Psi'' +\frac{1}{r}\Psi' - \Psi'^2 = -\frac{n^2}{r^2} .
\]
We, thus, find the remaining metric functions:
\begin{eqnarray}
\gamma(r) & = & m^2 \ln(r/r_0) \nonumber \\
\Psi(r) & = & n\ln(r/r_0) - \ln \left[ \frac{(r/r_0)^{2n} + \kappa}{(1+\kappa)} \right] ,
\end{eqnarray}
where the constant $n$ is related to $l$ and $m$ through the expression $n^2 = l^2 + m^2$. 

Therefore, the exterior metric is given by:
\begin{equation}
ds^2 =  \left( \frac{r}{r_0} \right)^{-2n} W^2(r) \left[
\left( \frac{r}{r_0}\right)^{2m^2} (-dt^2 +dr^2) + B^2r^2d\theta^2 \right] + 
\left( \frac{r}{r_0} \right)^{2n} 
\frac{1}{W^2(r)} dz^2 ,
\end{equation}
where
\[
W(r) \equiv \frac{(r/r_0)^{2n} + \kappa}{(1+\kappa)} .
\]
Besides, the solution for the scalar field $\phi(r)$ in the exterior region 
is given by equation (22). 
The integration constants $B,l,n,m$ will be fully determined after the 
introduction of the matter fields. In the particular case of Brans-Dicke, 
metric (24) belongs to a class 
of metrics corresponding to the {\em case 1} in Sen's paper \cite{sen}, 
with an appropriate adjustment in the parameters. 

\subsection{The Internal Solution and Matching:}

We start by considering the full modified Einstein's equations (14) 
with source (15) in the internal region defined by $ r \leq r_0$. In this region, 
all fields contribute to the energy-momentum tensor. In what follows, we will 
consider the solution for t
he superconducting string to linear order in $\tilde{G}_{0}$. 
Therefore, we assume that the metric $g_{\mu\nu}$ and the scalar field $\phi$ can be written as:
\[
g_{\mu\nu} = \eta_{\mu\nu} + h_{\mu\nu} ,
\]
\[
\phi = \phi_0 + \phi_{(1)} ,
\]
where $\eta_{\mu\nu} = diag(-,+,+,+)$ is the Minskowski metric tensor and $\phi_0$ is a constant. 
Thus, our problem reduces to solve the linearised Einstein's equations\footnote{To linear order in 
$\tilde{G}_{0}$, the modified Einstein's equations (10) reduce to the usual 
linearised Einstein's equations \cite{dam1}, the electromagnetic 
field being as in Minkowski spacetime.}
\begin{equation}
\nabla^2 h_{\mu\nu}= - 16\pi G\Omega^2(\phi_0)(T_{\mu\nu}^{(0)} - 
\frac{1}{2}\eta_{\mu\nu}T^{(0)}) ,
\end{equation}
in a harmonic coordinate system such that 
$(h^{\mu}_{\nu}-\frac{1}{2}\delta^{\mu}_{\nu}h)_{,\nu}=0$. 
$T^{(0)}_{\mu\nu}$ is the string's energy-momentum tensor to 
zeroth-order in $\tilde{G}_{0}=G\Omega^2(\phi_0)$ (evaluated in flat space) 
and $T^{(0)}$ its trace. Besides, we also need to solve the linearised equation for the scalar field
\begin{equation}
\nabla^2 \phi_{(1)}= - 4\pi G\Omega^2(\phi_0)\alpha(\phi_0)T^{(0)}.
\end{equation}
Then, we proceed with the junction between the internal and external solutions at 
$r= r_0$, with both solutions evaluated to linear order in $\tilde{G}_{0}$. 

While doing these calculations, we will briefly recall the method of
linearization using distribution functions (presented in Linet's paper 
\cite{lin} and applied later by Peter and Puy in \cite{pet2}, in the framework of General Relativity). 

\subsubsection{The Linearised Field Equations:}

First of all, let us evaluate the superconducting string's energy-momentum 
tensor to zeroth-order in $\tilde{G}_{0}$ in cartesian coordinates $(t,x,y,z)$. 
The non-vanishing components of the energy-momentum tensor can now be re-written under the form:
\begin{eqnarray}
T^{(0)t}_{t} & = & -\frac{1}{2} \left[ R'^2 + \sigma'^2 +\frac{R^2P^2}{r^2} + \sigma^2A^2  
 + (\frac{A'^2}{4\pi e^2}) + (\frac{P'^2}{4\pi q^2}) + 2V \right] \nonumber \\
T^{(0)x}_{x} & = & (\cos^2\theta -\frac{1}{2}) \left[ R'^2 + \sigma'^2 - 
\frac{R^2P^2}{r^2} + (\frac{A'^2}{4\pi e^2}) \right] 
 - \frac{1}{2} \left[ \sigma^2A^2 - (\frac{P'^2}{4\pi q^2}) + 2V \right] \nonumber \\
T^{(0)y}_{y} & = & (\sin^2\theta -\frac{1}{2}) \left[ R'^2 + \sigma'^2 -  
\frac{R^2P^2}{r^2} + (\frac{A'^2}{4\pi e^2}) \right] 
- \frac{1}{2} \left[ \sigma^2A^2 - (\frac{P'^2}{4\pi q^2}) + 2V \right] \nonumber \\
T^{(0)z}_{z} & = & -\frac{1}{2} \left[ R'^2 + \sigma'^2 +\frac{R^2P^2}{r^2} - \sigma^2A^2  
 - (\frac{A'^2}{4\pi e^2}) + (\frac{P'^2}{4\pi q^2})^2 + 2V \right]  
\end{eqnarray}

With the help of the source tensor defined by Thorne \cite{kip}, 
we can stablish some linear densities which will be very useful in our further analysis. Let 
\[
M^{\mu}_{\nu}(r) \equiv -2\pi \int_0^r T^{\mu}_{\nu}(r') r' dr'
\]
be the source tensor. Let us define its components (to zeroth-order in 
$\tilde{G}_{0}$) as follows. The energy per unit length $U$:
\[
U \equiv M^t_t = -2 \pi \int_0^{r_0} T^t_t r dr ;
\]
the tension per unit length $\tau$:
\[
\tau \equiv M^z_z = - 2\pi \int_0^{r_0} T^z_z r dr ;
\]
and the remaining transversal components as:
\begin{eqnarray*}
X & \equiv & M^r_r = -2\pi \int_0^{r_0} T^r_r r dr  \\
Y & \equiv & M^{\theta}_{\theta} = -2\pi \int_0^{r_0} T^{\theta}_{\theta} r dr .
\end{eqnarray*}
Now, in terms of the cartesian components of the energy-momentum tensor 
we can define a quantity $Z$ such that
\[
Z = -\int r dr d\theta T^x_x = - \int r dr d\theta T^y_y .
\]
If we assume that the string is (idealistically) infinetely thin, 
then its energy-momentum tensor may be described in terms of 
distribution functions. Namely, 
\begin{equation}
T^{\mu\nu} = diag (U, -Z, -Z, -\tau) \delta(x)\delta(y) .
\end{equation}
Equation (27) represents the string's energy-momentum tensor with all 
quantities integrated in the internal region $r \leq r_0$, in the 
cartesian coordinate system. Let us now evaluate the electromagnetic 
energy-momentum tensor (16) to zeroth-order in 
$\tilde{G}_{0}$ in cartesian coordinates. We can find easily that:
\begin{eqnarray}
T^{tt}_{em} & = & T^{zz}_{em} = \frac{I^2}{2\pi r^2} \nonumber \\
T^{ij}_{em} & = & \frac{I^2}{2\pi r^4} (2x^i x^j - r^2 \delta_{ij})
\end{eqnarray}
where $i,j = x,y$. Though the energy-momentum tensor (29) 
expresses the string's energy in the exterior region, one can still write 
it in terms of distributions, taking into account the relations
\[
\nabla^2 \left( \ln \frac{r}{r_0} \right)^2 = \frac{2}{r^2} \;\; \mbox{and} \;\; 
\partial_i\partial_j \ln \left( \frac{r}{r_0} \right) = \frac{(r^2\delta^{ij} - 2x^i x^j )}{r^4} .
\]
Therefore, (29) becomes
\begin{eqnarray}
T^{tt}_{em} & = & T^{zz}_{em} = \frac{I^2}{4\pi} \nabla^2 \left( 
\ln \frac{r}{r_0} \right)^2  \nonumber \\
T^{ij}_{em} & = & - \frac{I^2}{2\pi}  \partial_i\partial_j \ln \frac{r}{r_0} .
\end{eqnarray}
We are now in a position to calculate the linearised Einstein's equations (25) 
with source identified by:
\begin{eqnarray}
T^{tt}_{(0)} & = & U\delta(x)\delta(y) + \frac{I^2}{4\pi}\nabla^2 \left ( 
\ln \frac{r}{r_0} \right)^2 , \nonumber \\
T^{zz}_{(0)} & = & -\tau \delta(x)\delta(y) + \frac{I^2}{4\pi}\nabla^2 \left( 
\ln \frac{r}{r_0} \right)^2 , 
\nonumber \\
T^{ij}_{(0)} & = & I^2 \left[ \delta^{ij} \delta(x)\delta(y) - 
\frac{\partial_i\partial_j \ln \frac{r}{r_0}}{2\pi} \right] , 
\end{eqnarray}
and trace given by:
\begin{equation}
T_{(0)} = - (U +\tau - I^2 )\delta(x)\delta(y) .
\end{equation}
A straightforward calculus lead to the following solution of eq. (25):
\begin{eqnarray}
h_{00} & = & - 4\tilde{G}_{0} \left[ I^2 \ln^2 \frac{r}{r_0} + 
(U -\tau + I^2) \ln \frac{r}{r_0} \right] , \nonumber \\
h_{zz} & = & - 4\tilde{G}_{0} \left[ I^2 \ln^2 \frac{r}{r_0} + 
(U-\tau -I^2) \ln \frac{r}{r_0} \right] , \nonumber \\
h_{ij} & = & -4 \tilde{G}_{0} \left[ \frac{I^2}{2} r^2 \partial_i\partial_j 
\ln \frac{r}{r_0} + (U +\tau + I^2) \delta_{ij} \ln \frac{r}{r_0} \right] .
\end{eqnarray}
One can easily verify that the harmonic conditions 
$(h^{\mu}_{\nu} - \frac{1}{2}\delta^{\mu}_{\nu} h)_{,\nu} = 0$, 
with $h_{\mu\nu}$ given by (33), are identically satisfied. 
Using expression (32) for the trace of the energy-momentum tensor, we can solve 
eq. (26) straightforwardly:
\begin{equation}
\phi_{(1)} = 2\tilde{G}_{0} \alpha(\phi_0) (U+\tau -I^2)\ln \frac{r}{r_0} .
\end{equation}
As expected, since the linearised (modified) Eisntein's equations are 
the same as in General Relativity, we re-obtained here the same solutions 
(34) as in refs. \cite{lin,pet2}. However, the scalar-tensor feature 
still brings a new information coming from solution (34). 

We return now to the original cylindrical coordinates system and obtain:
\begin{eqnarray}
g_{tt} & = & - \left\{ 1 + 4\tilde{G}_{0} \left[ I^2 \ln^2 \frac{r}{r_0} + 
(U-\tau +I^2)\ln \frac{r}{r_0} \right]\right\} , \nonumber \\
g_{zz} & = & 1 - 4\tilde{G}_{0} \left[ I^2 \ln^2 \frac{r}{r_0}
+ (U-\tau - I^2)\ln \frac{r}{r_0} \right] , \nonumber \\
g_{rr} & = & 1 + 2\tilde{G}_{0}I^2 - 4\tilde{G}_{0}(U +\tau + I^2) \ln \frac{r}{r_0} , \nonumber \\
g_{\theta\theta} & = & r^2 \left[ 1 - 2\tilde{G}_{0}I^2 -
4\tilde{G}_{0} (U+\tau + I^2) \ln \frac{r}{r_0} \right] .
\end{eqnarray}

In order to preserve our previous assumption that $g_{tt}=-g_{\rho\rho}$ 
(corresponding to the particular case of a magnetic solution of the 
Einstein-Maxwell-dilaton eqs.), we make a change of variable $r \rightarrow \rho$, such that
\[
\rho = r \left[ 1 + \tilde{G}_{0} (4U + I^2) - 4\tilde{G}_{0} U \ln \frac{r}{r_0} - 
2\tilde{G}_{0}I^2 \ln^2 \frac{r}{r_0}\right] ,
\]
and, thus, we have
\begin{eqnarray}
ds^2 & = & \left\{ 1 + 4\tilde{G}_0 \left[ I^2\ln^2 \frac{\rho}{r_0} + (U-\tau +I^2)\ln 
\frac{\rho}{r_0}\right] \right\} (-dt^2 + d\rho^2) \nonumber \\
& & + \left\{ 1 - 4\tilde{G}_0 \left[ I^2 \ln^2\frac{\rho}{r_0} +
(U-\tau -I^2) \ln \frac{\rho}{r_0} \right]\right\} dz^2  \\
& & + \rho^2 \left[ 1 - 8\tilde{G}_0 (U+ \frac{I^2}{2}) + 
4\tilde{G}_0 (U-\tau -I^2)\ln \frac{\rho}{r_0} + 4\tilde{G}_0 I^2 
\ln^2\frac{\rho}{r_0} \right] d\theta^2 \nonumber . 
\end{eqnarray}
Expressions (34) and (36) represent, respectively, the solutions of the scalar field and an 
isolated current-carrying string in the conformal frame, as long as the weak-field 
approximation is valid. 
Comparison with the external solutions (22) and (24) requires a 
linearision of these ones since they are exact solutions. 
Expanding them in power series of the paramenters $m$ and $n$, we find 
\begin{eqnarray*}
g_{\rho\rho} & = & - g{tt} = 1 + 2m^2 \ln \frac{\rho}{r_0} + h(\rho) \\
g_{zz} & = & \frac{1}{1+h(\rho)} \\
g_{\theta\theta} & = & B^2 \rho^2 [ 1 + h(\rho) ]  ,
\end{eqnarray*}
with
\[
h(\rho) = 2n \frac{1 - \kappa}{1+\kappa} \ln \frac{\rho}{r_0} + 
2n^2 \frac{1 + \kappa^2}{(1+\kappa)^2} \ln^2 \frac{\rho}{r_0}  .
\]
Making the identification of the coefficients of both linearised metrics, we finally obtain
\begin{eqnarray}
m^2 & = & 4\tilde{G}_0 I^2 \nonumber \\
B^2 & = & 1 - 8\tilde{G}_0 (U+\frac{I^2}{2}) \nonumber \\
l & = & 2\tilde{G}_0 \alpha(\phi_0) (U+\tau-I^2) \nonumber \\
\kappa & = & 1 + \tilde{G}^{1/2}_0 (U - \tau -I^2) .
\end{eqnarray}
Calculating now the deficit angle for metric (36)
\[
\Delta\theta = 2\pi \left[ 1 - \frac{1}{\sqrt{g_{\rho\rho}}}\frac{d}{d\rho} 
\sqrt{g_{\theta\theta}} \right] ,
\]
we finally obtain
\begin{equation}
\Delta\theta = 4\pi \tilde{G}_0 (U +\tau + I^2) .
\end{equation}

\subsection{Bending of Light Rays:}

A light ray coming from infinity in the transverse plane has its trajectory 
deflected, for an observer at infinity,  by an angle given by:
\[
\Delta\theta = 2 \int_{\rho_{min}}^{\infty} d\rho 
[-\frac{g^2_{\theta\theta}p^{-2}}{g_{\rho\rho}g_{tt}} - 
\frac{g_{\theta\theta}}{g_{\rho\rho}}]^{-1/2} \,\, - \pi 
\]
where $\rho_{min}$ is the distance of closest approach, given by 
$\frac{d\rho}{d\theta} =0$:
\[
\frac{g_{\theta\theta}(\rho_{min})}{g_{tt}(\rho_{min})} = -p^2
\]
which gives in turn:
\[
\frac{\rho_{min}}{r_0} = (\frac{p}{Br_0})^{1/(1-m^2)}.
\]

We can now evaluate the deficit angle to first order in $\tilde{G}_0$. 
Performing an expansion to linear order in this factor, 
in much the same way as Peter and Puy \cite{pet2}, we find:
\[
\Delta\theta = \frac{2}{B(1-m^2)}[\frac{\pi}{2}(1+m^2\ln\frac{p}{Br_0})-m^2\nu]
-\pi ,
\]
where we have defined the quantity $\nu$ as 
\[
\nu \equiv - \int_0^1\frac{\ln s}{\sqrt{1-s^2}} ds = \frac{\pi}{2}\ln 2 , 
\]
with $s \equiv \frac{p}{Br_0}(\frac{\rho}{r_0})^{m^2 -1}$. Using expressions (37), we 
have
\begin{equation}
\Delta\theta = 4\pi \tilde{G}_0 \left[ U + I^2 \left(\frac{3}{2} + 
\ln \frac{\rho}{r_0}\right)\right] + 8\nu \tilde{G}_0 I^2 .
\end{equation}

\section{Conclusion}

In this work we studied the modifications induced by a scalar-tensor gravity 
on the metric of a current-carrying string described by model with action 
given by eq. (2). For this purpose, 
we made an adaptation of Linet's method which consists in linearising the 
Einstein's and dilaton's equations using distribution's functions while taking into account 
the scalar-tensor feature of gravity. We found that the metric depends on five 
parameters which are related to the string's internal structure and to the 
scalar field (dilaton) solution. 
Concerning the deflection of light, if we compare our results with those 
obtained in General Relativity, we see that expression (39) does not change 
substantially, albeit the metric structure is indeed modified with respect to 
the one in General Relativity.  

Now, an interesting investigation that opens up, and we have already 
initiated to pursue, is the analysis of the properties of the cosmic 
string generated by the action (2) in the supersymmetrized version, 
where it is implicit that the scalar-tensor degrees of freedom 
of the gravity sector are accomodated in a suitable supergravity multiplet. The study 
of such a model raises the question of understanding the r\^ole played by the 
fermionic partners of the bosonic matter and by the gravitino in the configuration of a string. 
Also, it might be of relevance to analyse the possibility of 
gaugino and gravitino condensation in this scenario. 

\section*{Acknowledgements}

The authors are grateful to Brandon Carter, Bernard Linet and Patrick Peter for 
many discussions, suggestions and a critical reading of this manuscript. 
One of the authors (MEXG) thanks to the Centro Brasileiro de 
Pesquisas F\'{\i}sicas (in particular, the Departamento de Campos e Part\'{\i}culas) 
and to the Abdus Salam ICTP-Trieste 
for hospitality during the preparation of part of this work. CNF thanks to CNPq for a PhD grant.


\begin{thebibliography}{99}
\bibitem{dam} Th. Damour, {\em gr-qc/9904057 ; Nuclear Phys. B (to appear)} .
\bibitem{green} M. B. Green, J. H. Schwarz and E. Witten, 
{\em Superstring Theory} (Cambridge: Cambridge Univ. Press, 1987).
\bibitem{rom} C. Gundlach and M. E. Ortiz, {\em Phys. Rev. D} {\bf 42} (1990), 2521; 
L. O. Pimentel and A. No\'e Morales, {\em Revista Mexicana de F\'{\i}sica} {\bf 36} (1990), S199; 
A. Barros and C. Romero, {\em J. Math. Phys.} {\bf 36} (1990), 5800.
\bibitem{greg} R. Gregory and C. santos, {\em Phys. Rev. D.} {\bf 56} (1997), 1194.
\bibitem{mexg} M. E. X. Guimar\~aes, {\em Class. Quantum Gravity} {\bf 14} (1997), 435.
\bibitem{wit} E. Witten, {\em Nuclear Phys.} {\bf B 249} (1995), 557.
\bibitem{tan} T. Vachaspati, {\em Phys. Lett. B} {\bf 265} (1991), 258; 
R. H. Brandenberger et alli., {\em Phys. Lett. B} {\bf 293} (1992), 287; 
K. Dimopoulos, {\em Phys. Rev. D} {\bf 57} (1998), 4629.
\bibitem{nas} J. R. S. Nascimento, I. Cho and A. Vilenkin, {\em hep-th/9902135.}
\bibitem{hill} C. T. Hill, D. N. Schramm and T. P. Walker, {\em Phys. Rev. D} {\bf 36} (1987), 1007.
\bibitem{sen} A. A. Sen, {\em Phys. Rev. D} {\bf 60} (1999), 067501.
\bibitem{moss} I. Moss and S. Poletti, {\em Phys. Lett. B} {\bf 199} (1987), 34.
\bibitem{ams} P. Amsterdanski and P. Laguna-Castillo, {\em Phys. Rev. D} {\bf 37} (1988), 877.
\bibitem{babul} A. Babul, T. Piran and D. N. Spergel, {\em Phys. Lett. B} {\bf 209} (1988), 477.
\bibitem{lin} B. Linet, {\em Class. Quantum Gravity} {\bf 6} (1989), 435.
\bibitem{helli} T. M. Helliwell and D. Konkowski, {\em Phys. Lett. A} {\bf 143} (1990), 438.
\bibitem{pet2} P. Peter and D. Puy, {\em Phys. Rev. D} {\bf 48} (1993), 5546.
\bibitem{niel} H. B. Nielsen and P. Olesen, {\em Nucl. Phys.} {\bf B 61} (1973), 45.
\bibitem{louis} L. Witten, {\em Gravitation: An Introduction to Current Research} 
(ed. L. Witten, New York: Wiley, 1962).
\bibitem{dam1} T. Damour and K. Nordtverdt, {\em Phys. Rev. D} {\bf 48} (1993), 3436.
\bibitem{kip} K. S. Thorne, {\em Phys. Rev.}{\bf 138} (1965), 251. 








\end{thebibliography}
\end{document}